\documentclass[conference]{IEEEtran}

\usepackage[cmex10]{amsmath}
\usepackage{amssymb}  
\usepackage{siunitx}  
\usepackage[capitalise]{cleveref} 
    \crefformat{equation}{(#2#1#3)}	

\ifCLASSOPTIONcompsoc  \usepackage[nocompress]{cite}
\else                  \usepackage{cite}
\fi

\usepackage{url}

\usepackage{graphicx}

\ifCLASSOPTIONcompsoc	\usepackage[caption=false,font=normalsize,labelfont=sf,textfont=sf]{subfig}
\else \usepackage[caption=false,font=footnotesize]{subfig}
\fi

\graphicspath{{Figures/}}

\begin{document}

\title{Traffic-Aware UAV Placement using a Generalizable Deep Reinforcement Learning Methodology}

\author{
    \IEEEauthorblockN{
        Eduardo Nuno Almeida,
        Rui Campos,
        Manuel Ricardo
    }
    \IEEEauthorblockA{
        INESC TEC and Faculdade de Engenharia, Universidade do Porto, Portugal \\
        \{eduardo.n.almeida, rui.l.campos, mricardo\}@inesctec.pt
    }
}

\maketitle

\begin{abstract}

Unmanned Aerial Vehicles (UAVs) acting as Flying Access Points (FAPs) are being used to provide on-demand wireless connectivity in extreme scenarios. Despite ongoing research, the optimization of UAVs' positions according to dynamic users' traffic demands remains challenging.
We propose the Traffic-aware UAV Placement Algorithm (TUPA), which positions a UAV acting as FAP according to the users' traffic demands, in order to maximize the network utility. Using a DRL approach enables the FAP to autonomously learn and adapt to dynamic conditions and requirements of networking scenarios. Moreover, the proposed DRL methodology allows TUPA to generalize knowledge acquired during training to unknown combinations of users' positions and traffic demands, with no additional training.
TUPA is trained and evaluated using network simulator ns-3 and ns3-gym framework. The results demonstrate that TUPA increases the network utility, compared to baseline solutions, increasing the average network utility up to 4x in scenarios with heterogeneous traffic demands.

\end{abstract}


\section{Introduction} \label{Introduction-Section}


\IEEEPARstart{T}{he} use of Unmanned Aerial Vehicles (UAVs) acting as aerial Wi-Fi Access Points (APs) or cellular base stations has emerged as an interesting solution to provide on-demand wireless connectivity. They have been considered to provide or extend the coverage of existing network infrastructures or to enhance their capacity and accommodate temporary traffic demand surges \cite{mozaffari2019tutorial}. Aerial networks are especially interesting in emergency and crowded scenarios, where the unpredictability and heterogeneity of the users' mobility and traffic demands may lead to the inefficient utilization of the network's radio resources. In these heterogeneous scenarios, UAVs can be dynamically positioned according to the users' traffic demands, in order to meet their Quality of Service (QoS) requirements. Despite the ongoing research on UAV placement \cite{almeida2021joint, wu2018common, liu2019trajectory, lai2019demand}, some challenges remain open, including the optimization of the UAVs' positions according to the dynamic users' traffic demands.


Reinforcement learning (RL), including Deep RL (DRL), is being used to solve traditional wireless networks problems \cite{feriani2021single}, including UAV placement \cite{tang2020minimum, anokye2021deep, xie2021connectivity}. RL allows an agent to autonomously learn the optimal policy of actions that solve a problem, using the experience collected in previous interactions with the environment. This enables the agent to learn and adapt to the current scenario dynamics, including the mobility of the users and their traffic demands, instead of relying on predefined models, which may not accurately characterize the scenario. Moreover, the agent can be trained with real-time network metrics, in order to learn and react to the real conditions of the environment, instead of using approximations or estimations that may be inaccurate or delayed.


Generally, RL problems are designed to solve a specific scenario with predefined parameters and conditions. Therefore, the RL agent is trained and evaluated in the same exact scenario and conditions defined in the problem, in order to achieve maximum performance. However, this methodology results in an RL policy that may not be sufficiently generalizable. In fact, when the resulting RL policy is applied in similar but unknown scenarios, with different parameters and conditions, its performance may significantly degrade \cite{cobbe2019quantifying}. One solution to overcome this problem is to adopt an online training methodology, where the RL agent is continuously refining its policy over time, in order to adapt to new and unpredictable environment conditions that may occur in the future. Still, the time needed to learn the new policy and the increased complexity overhead limit the efficiency of this methodology.


The main contribution of this paper is the \textbf{Traffic-aware UAV Placement Algorithm (TUPA) using a generalizable DRL methodology}. The objective of TUPA is to position a UAV acting as a Flying Access Point (FAP) according to the users' traffic demands, thus maximizing the network utility provided to the users. The use of a DRL approach techniques enables the FAP to autonomously learn and adapt to the dynamic conditions and requirements of different scenarios, in order to efficiently manage and provide network capacity to the users that actually require it. Moreover, the proposed DRL methodology allows TUPA to generalize the knowledge acquired in the training phase to unknown variations of scenarios without requiring additional training. In this paper, we define a scenario as a combination of the users' positions and their generated traffic. Additionally, we use the terms UAV and FAP interchangeably.

In order to evaluate TUPA, including its generalization capacity and learning rate, we compare two offline training strategies. In the \textbf{specialization strategy}, the FAP is trained in the same scenarios used in the evaluation; in the \textbf{generalization strategy}, the FAP is trained in similar but different scenarios. In both strategies, TUPA is evaluated in the same exact scenarios and conditions. The comparison of both strategies allows the study of the effect of the training set on the overall performance of TUPA, as well as its generalization capacity.

TUPA is trained and evaluated using the ns3-gym framework \cite{gawowicz2019ns3}, which enables the development of an OpenAI Gym RL environment based on an underlying ns-3 simulation \cite{ns3Simulator}. Using ns-3 enables TUPA to be trained and evaluated with a much larger number of scenarios, which would be very difficult to achieve in an experimental testbed. The training and evaluation is driven by real-time network metrics of the wireless environment, which are calculated with accurate wireless network models in ns-3. Still, ns-3 can also be extended with the trace-based simulation approach to replicate the same exact conditions observed in past experimental scenarios \cite{fontes2017trace}.


The rest of this paper is organized as follows.
\cref{SystemModel-Section} presents the system model and the problem formulation.
\cref{TUPA-Section} explains the proposed TUPA.
\cref{Results-Section} discusses the training and performance evaluation methodologies and analyzes the simulation results.
\cref{Conclusions-Section} presents the conclusions and future work.

\section{System Model and Problem Formulation} \label{SystemModel-Section}

\begin{figure}
    \centering
    \includegraphics[width=0.6\linewidth]{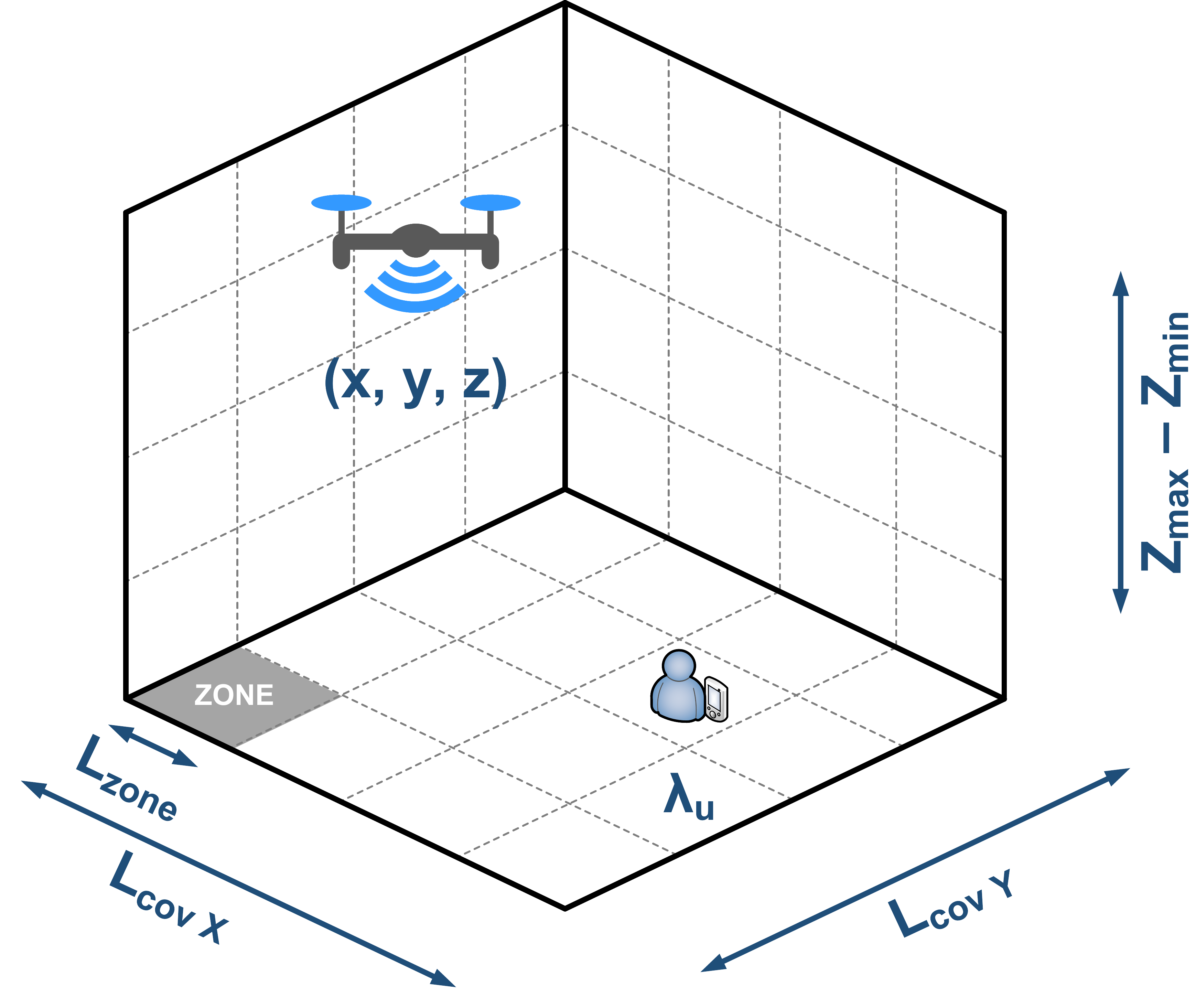}
    \caption{System model representing the coverage area, the limits for the UAV altitude and the position of the users.}
    \label{SystemModel-Figure: System model}
\end{figure}

\cref{SystemModel-Figure: System model} represents the model of the system. Let the area to be covered by the FAP be represented as a rectangle $ L_{\textrm{cov}_\textrm{X}} \times L_{\textrm{cov}_\textrm{Y}} $. The coverage area is subdivided in squares $ L_{\textrm{zone}} \times L_{\textrm{zone}} $, defining zones, which represent and aggregate all users within that subarea. The wireless network is composed of 1 FAP and $ U $ users positioned within the coverage area. The FAP can only be positioned within the volume $ [0, L_{\textrm{cov}_\textrm{X}}] \times [0, L_{\textrm{cov}_\textrm{Y}}] \times [Z_{\textrm{min}}, Z_{\textrm{max}}] $. Each user generates UDP constant bit-rate (CBR) traffic of $ \lambda_u $ \si{bit/s}. Regardless of the individual values, the aggregate traffic generated by all users is always the same, which is given by $ \Lambda = \sum_{u=1}^{U} \lambda_u $.

In this paper, we define a \emph{scenario} as a fixed wireless environment with a given combination of users' positions and their generated traffic. Using the standard RL terminology, we define an \emph{episode} as a sequence of $ T $ time steps taken in a given \emph{scenario}. Each time step $ t $ has a duration of $ S $ seconds. During the episode, the positions of the users and their generated traffic remain constant. At the end of time step $ t $, the network utility $ N[t] $ is calculated as a linear combination of the average aggregate throughput $ T[t] $, average delay $ D[t] $ and average Packet Loss Ratio (PLR) $ P[t] $, calculated among all users, as defined in \cref{SystemModel-Equation: Network utility}.

\begin{equation} \label{SystemModel-Equation: Network utility}
    N[t] = \alpha_T \cdot T[t] + \alpha_D \cdot (1 - D[t]) + \alpha_P \cdot (1 - P[t])
\end{equation}

The parameters $ \alpha_T $, $ \alpha_D $ and $ \alpha_P $ are the relative weights of the throughput, delay and PLR, respectively. All three components are normalized in order for the values to be approximately in the interval $ [0, 1] $. The throughput is normalized using the aggregate throughput $ \Lambda $ generated by the users. The delay does not have a maximum known value; therefore, it is normalized using an empirical value so that most delays can be normalized in the interval $ [0, 1] $. In corner cases, time steps may have average delays slightly greater than 1, which do not affect the overall training process nor the evaluation of TUPA. The PLR is already a normalized value.

Therefore, for a given episode $ e $, the objective is to determine a sequence of $ T $ FAP positions that maximizes the episode's average network utility $ N_e $, as stated in \cref{SystemModel-Equation: Objective function}.

\begin{equation} \label{SystemModel-Equation: Objective function}
    \max N_e = \frac{1}{T} \sum_{t=1}^{T} N[t]
\end{equation}

\section{Traffic-Aware DRL UAV Placement Algorithm} \label{TUPA-Section}

\begin{figure}
    \centering
    \includegraphics[width=1\linewidth]{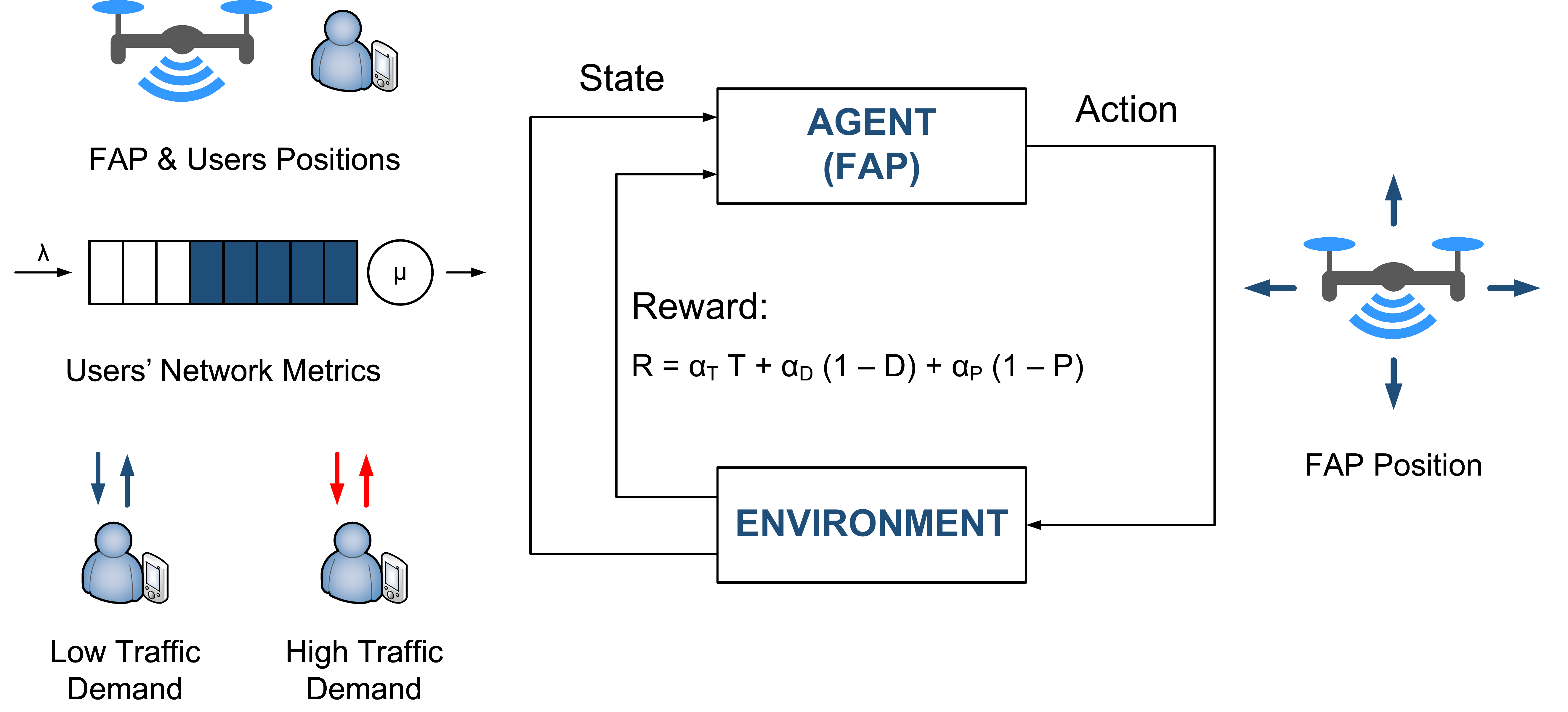}
    \caption{Proposed traffic-aware DRL UAV placement algorithm TUPA, depicting the observations, actions and reward function.}
    \label{TUPA-Figure: Traffic-aware DRL FAP placement}
\end{figure}

The traffic-aware DRL UAV placement algorithm TUPA is depicted in \cref{TUPA-Figure: Traffic-aware DRL FAP placement}. The problem is formulated as a standard RL problem. The UAV acting as a FAP is the agent that should learn the optimal policy of actions to take on the environment for any given state, which is represented by a set of observations collected from the environment. As a result of the action taken on a given state, the environment will reward the agent with a value proportional to the quality of the action.

As in standard RL problems, the FAP agent is trained and evaluated in multiple episodes. As defined in \cref{SystemModel-Section}, an \emph{episode} is a sequence of $ T $ time steps taken on a \emph{scenario}. In each time step $ t $, the agent collects the observations $ O[t] $ from the environment, determines the action $ A[t] $ to take on that state and receives a reward $ R[t] $ from the environment. An episode finishes after $ T $ time steps. Then, the episode results are calculated considering all time steps of the episode, such as the cumulative reward, average throughput, average delay, average PLR and the resulting average network utility. Therefore, the objective is to train the FAP agent through multiple episodes of different scenarios, so that it learns, over time, the optimal policy of actions to take on the environment that maximizes the episode's cumulative reward.

The following sections explain the environment model, including the observations, actions and the reward function, as well as the DRL algorithms used to train the FAP agent.

\subsection{Environment's Observation Space}

The environment's state for a given time step is represented by a set of observations collected from the environment, which are related to the state of the network and the users' traffic demands in that time step.

Each observation is a 2-D matrix representing the zones of the coverage area defined in \cref{SystemModel-Section}. Hence, the dimensions of each matrix are $ ( L_{\textrm{cov}_\textrm{X}} / L_{\textrm{zone}} ) \times ( L_{\textrm{cov}_\textrm{Y}} / L_{\textrm{zone}} ) $. The set of observations $ O[t] $ for time step $ t $ are as follows: 1) the distance between the FAP and each zone's center; 2) the number of users in each zone; 3) the aggregate traffic offered by the users in each zone; 4) the aggregate throughput (transported traffic) of the users in each zone; 5) the average delay of the users in each zone; and 6) the average PLR of the users in each zone.

As defined in \cref{SystemModel-Section}, every observation is normalized. The normalization is performed using the maximum theoretical value of that observation. Moreover, all observation matrices $ O_i $, except the distance, take the complementary value $ (1 - O_i) $, so that the matrices are dense (i.e., with a majority of non-zero values).

\subsection{Agent's Action Space} \label{TUPA-Section: Action space}

In order to evaluate and compare alternative approaches to TUPA, we define two distinct action spaces for the agent: 1) sequential movements, 2) absolute coordinates.

In the sequential movements discrete action space, the UAV moves through a sequence of predefined movements, once per time step. In each time step, the UAV can only move in one of 6 directions: \{Up, Down, Left, Right, Front, Back\}. For each direction, the UAV can move 1 or 5 zones at once. This enables the UAV to quickly move towards the desired area, while being able to refine its position. Alternatively, the UAV can choose to stand in the current position. Despite the 13 actions defined in this action space, only the actions that lead the UAV to a valid position, within the boundaries defined in \cref{SystemModel-Section}, are selectable in that time step.

The absolute coordinates continuous action space considers that the UAV can select, in each time step, the absolute coordinates of its next position, defined by the tuple $ (x, y, z) $, within the valid volume. Similarly to the observations, the UAV coordinates $ (x, y, z) $ are normalized to the interval $ [0, 1] $, which represent the minimum and maximum limits of the coverage area.

\subsection{Reward Function}

In order to drive the training of the agent towards the objective defined in \cref{SystemModel-Equation: Objective function}, the reward function is defined as the network utility of the time step, which is given by \cref{SystemModel-Equation: Network utility}. Thus, the reward $ R[t] $ calculated for time step $ t $ is given by the network utility $ N[t] $ of that time step.

\subsection{Deep Reinforcement Learning Algorithms}

As explained in \cref{TUPA-Section: Action space}, two action spaces are considered for the agent: the sequential movements (discrete) and the absolute coordinates (continuous). Since the two action spaces have different dimensions, each action space requires a different DRL algorithm to train the agent, according to the action space's dimensions.

The Double Deep Q-Network (DDQN) algorithm is used to train the agent using the sequential movements discrete action space, whereas the Deep Deterministic Policy Gradient (DDPG) algorithm is used for the absolute coordinates continuous action space.

The algorithms are implemented with neural networks composed of 1-2 convolutional layers with kernels of size (3,3), followed by 2 dense layers interleaved with dropout layers. All layers use the Rectified Linear Unit (ReLU) as the activation function. The RMSprop optimizer with a learning rate between $ 10^{-3} $ and $ 10^{-4} $ is used in both algorithms, associated to the mean squared error loss function for the DDQN algorithm and the Huber loss function for the DDPG algorithm. The discount factor $ \gamma $ equals 0.9. The DDQN algorithm uses an $ \epsilon $-greedy exploration strategy, whose $ \epsilon $ starts with an initial value of 1 and decays exponentially along up to 1000 episodes with a decay rate of 0.02. Finally, the agent is trained after each time step. In order to stabilize the training, the DDQN and the DDPG algorithms use target networks that are 80\% soft-updated after 1000 or 2000 steps, respectively. Moreover, a replay buffer with a capacity of 100,000 steps and a batch size of 64 steps is used in both algorithms.

\section{Simulation Results} \label{Results-Section}

The training and test methodologies are explained in this section, as well as the performance evaluation of TUPA by means of simulation.

\subsection{Training and Performance Evaluation Methodology}

TUPA was trained and evaluated using the ns3-gym framework \cite{gawowicz2019ns3}. ns3-gym enables the development of an OpenAI Gym RL environment based on an underlying ns-3 network simulation \cite{ns3Simulator}, which defines and runs the dynamic wireless networking environment where the agent is trained and evaluated. In this sense, the actions taken by the agent are passed to the ns-3 simulation environment, and the observations and rewards are calculated using the real-time network metrics collected by ns-3. The DDQN and DDPG DRL algorithms used in this work were implemented with the TF-Agents Python library \cite{tfAgents}.

\cref{Results-Table: Training and test parameters} summarizes the system model, ns-3.34 simulator and DRL parameters. The scenarios are setup according to the system model presented in \cref{SystemModel-Section}. The wireless network is configured in AP--STA infrastructure mode. The FAP--User wireless link is modeled by the Friis path loss and Rician fast-fading, according to the experimental channel model for UAVs hovering at low altitudes measured in \cite{almeida2021joint}. Each \emph{scenario} is characterized by a random combination of users' positions and offered loads, whose sum is always equal to $ \Lambda = \SI{40}{Mbit/s} $ generated at the application layer. Due to technical limitations of the ns3-gym framework, only 3 users were simulated. Nevertheless, these users are sufficient to generate a large number of different combinations of traffic demand composing the training and test datasets. Moreover, each user may approximately model a set of users generating the same aggregate traffic in a given geographic area. In order to evaluate the ideal performance of TUPA, the UAV is always positioned at the intended coordinates at the start of each time step $ t $. The effect of the UAV movement on TUPA's performance is left for future work. Finally, the relative weights of the average network utility $ \alpha_T = 0.6 $, $ \alpha_D = 0.3 $ and $ \alpha_P = 0.1 $ were selected as a trade-off between maximizing the throughput and delay, while ensuring all users remain connected to the FAP and do not starve.

\begin{table}
    \centering
    \caption{System Model, ns-3 Simulator and DRL Parameters}
    \label{Results-Table: Training and test parameters}
    \begin{tabular}{l l}
        \hline
        \multicolumn{2}{c}{\textbf{System Model Parameters}} \\
        \hline
        $ L_{\textrm{cov}_\textrm{X}} \times  L_{\textrm{cov}_\textrm{Y}}$ & \SI{500}{m} $ \times $ \SI{500}{m} \\
        $ L_{\textrm{zone}} $ & \SI{25}{m} \\
        $ [ Z_{\textrm{min}} $, $ Z_{\textrm{max}} ] $ & [\SI{25}{m}, \SI{100}{m}] \\
        $ U $ & 3 Users \\
        $ \Lambda $ & \SI{40}{Mbit/s} \\
        \hline
        \multicolumn{2}{c}{\textbf{ns-3.34 Simulator Parameters}} \\
        \hline
        Wi-Fi standard        & IEEE 802.11ac \\
        Wi-Fi channel         & 36 \\
        Channel bandwidth     & \SI{20}{MHz} \\
        Tx power              & \SI{20}{dBm} \\
        Propagation model     & Friis + Rician (K = \SI{13}{dB}) \\
        Minimum preamble RSSI & \SI{-90}{dBm} \\
        Application traffic   & UDP constant bitrate \\
        Packet length         & \SI{512}{Bytes} \\
        MAC queues            & 500 Packets \\
        MAC auto rate         & Ideal \\
        \hline
        \multicolumn{2}{c}{\textbf{DRL Parameters}} \\
        \hline
        $ T $ & 100 Time steps \\
        $ S $ & \SI{1}{s} \\
        \# Unique training scenarios
            & Generalization strategy: 10,000 or 500 \\
            & Specialization strategy: 500 \\
        \# Unique test scenarios & 500 \\
        \# Training episodes & 10,000 \\
        \# Test episodes & 500 \\
        $ \{ \alpha_T , \alpha_D , \alpha_P \} $ & \{0.6, 0.3, 0.1\} \\
        \hline
    \end{tabular}
\end{table}

In order to generate different levels of traffic demand heterogeneity, the users' traffic demands follow a random exponential distribution. Similarly, the users' $ x $ and the $ y $ coordinates also follow a random exponential distribution. This method allows the generation of homogeneous scenarios, where the users are close to each other and/or have similar traffic demands, and heterogeneous scenarios, where the users are away from each other and/or have different traffic demands.

In order to evaluate the performance of TUPA, including its generalization capacity, we defined two training strategies: 1) specialization and 2) generalization. In both strategies, we defined two sets of scenarios: a training set, composed of the scenarios where the FAP agent was trained; and the test set, composed of the scenarios to evaluate the FAP agent. The training set contains up to 10,000 unique scenarios, whereas the test set contains 500 unique scenarios. In both strategies, the FAP agent is always trained for 10,000 episodes. The comparison of both strategies allows the study of the effect of the training set on the overall performance of TUPA, as well as its generalization capacity.

The objective of the \textbf{specialization strategy} is to analyze the performance of TUPA when it is trained in the same exact scenarios as those used in the evaluation -- i.e., the training set equals the test set. Since the test set, and consequently the training set, have less unique scenarios than the number of training episodes, the FAP agent was trained multiple times per scenario (i.e., multiple episodes of the same scenarios). The FAP agent follows the given order: episode 1 is run on scenario 1, episode 2 is run on scenario 2, and so on. When the FAP agent finishes the last scenario of the training set, it returns to scenario 1 and continues the training by repeating that scenario again. This training process finishes once the FAP agent completes the specified number of training episodes.

In the \textbf{generalization strategy}, the FAP agent was trained in scenarios different from those used in the test set -- i.e., the training set is different from the test set. This strategy allows the analysis of TUPA's capacity to generalize the knowledge learned in different, but similar, scenarios and apply it to unknown scenarios, without additional training. Unlike the specialization strategy, in order to maximize the generalization capacity, the FAP agent should be trained with a much larger number of unique scenarios, instead of training in the same scenarios for multiple episodes \cite{cobbe2019quantifying}. In order to analyze the effect of the number of unique scenarios in the training set on the generalization capacity, we compare two versions of the generalization strategy: 1) training in 10,000 unique scenarios during one episode per scenario (\emph{Gen10k}); and 2) 500 unique scenarios during 20 episodes per scenario (\emph{Gen500}).

In addition to the network utility performance, we analyzed the learning rate of TUPA. To that end, the FAP agent was trained with a different number of training episodes and its performance was evaluated after each training session.

Therefore, we trained and tested 4 versions of TUPA: 1-2) the DDQN with the sequential movements action space using the specialization and generalization strategies, respectively; and 3-4) the DDPG with the absolute coordinates action space using the specialization and generalization strategies, respectively. Then, we compared the results obtained with a common baseline. In the baseline, the FAP was positioned in order to minimize the average distance to the users, which is equivalent to maximizing the average Signal-to-Noise Ratio (SNR) of the users. This baseline represents the state of the art solutions that are not traffic-aware and optimize the SNR of all users independently of the users' actual traffic demands. All versions of TUPA and the baseline were evaluated on the same set of test scenarios.

\subsection{Performance Evaluation}

In order to evaluate the performance of TUPA and analyze its generalization capacity, we trained the FAP agent with 10,000 episodes using the generalization (\emph{Gen10k} and \emph{Gen500}) and specialization (\emph{Spec}) strategies. After the training, the performance of TUPA was evaluated in the test scenarios. For each episode, we calculated the average network utility and a heterogeneity index of that scenario.

We define the heterogeneity index $ H_s $ of scenario $ s $ as a metric that characterizes the spatial heterogeneity of the users' traffic demands in scenario $ s $. The heterogeneity index $ H_s \in [0, 1] $ is given by \cref{Results-Equation: Heterogeneity index} and it is calculated as the average between two components: 1) the complementary value of the Jain's fairness index for the users' traffic demands $ \lambda_u $; and 2) the average normalized distance $ d_{ij} $ among all $ U $ users. Hence, a heterogeneity index close to 1 represents a scenario where the users have very heterogeneous traffic demands and/or are scattered throughout the coverage area; a heterogeneity index close to 0 represents the opposite (homogeneous scenario).

\begin{equation} \label{Results-Equation: Heterogeneity index}
    H_s =
    \frac{1}{2} \times \left[ 1 - \frac{ \left( \sum_{u=1}^{U} \lambda_u \right) ^ 2 }{ U \sum_{u=1}^{U} \lambda_u^2 } \right]
    +
    \frac{1}{2} \times \frac{1}{U (U-1)} \sum_{i=0}^{U} \sum_{\substack{j=0 \\ j \neq i}}^{U} d_{ij}
\end{equation}

\begin{figure}
    \centering
    \includegraphics[width=1\linewidth]{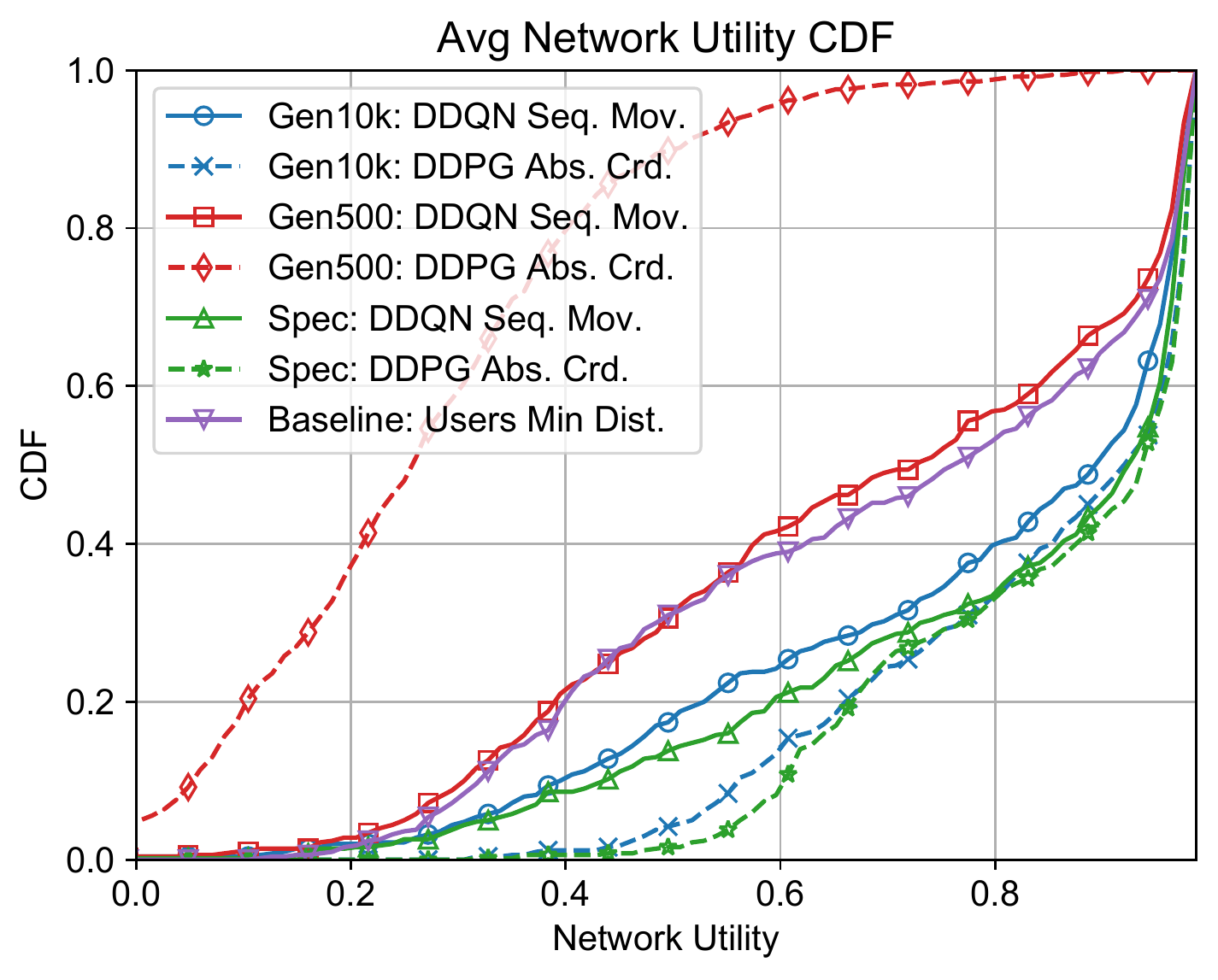}
    \caption{Average network utility calculated for all test episodes. All versions of TUPA were trained with 10,000 episodes.}
    \label{Results-Figure: Network utility CDF}
\end{figure}

\begin{figure*}
    \centering
    \subfloat[Average aggregate throughput CDF.] {
        \includegraphics[width=0.31\linewidth]{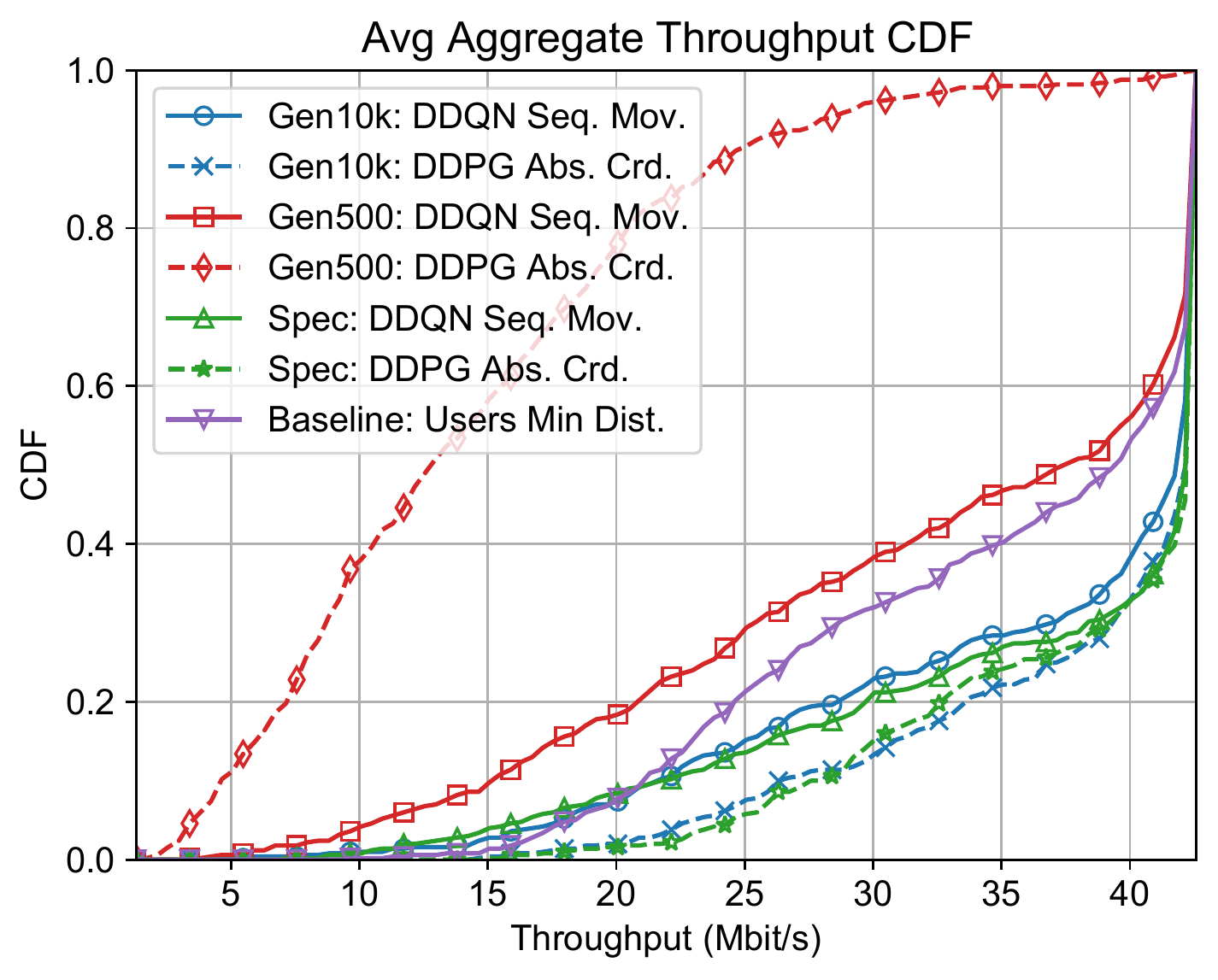}
        \label{Results-Figure: Throughput CDF}
    }
    \hfil
    \subfloat[Average delay CDF.] {
        \includegraphics[width=0.31\linewidth]{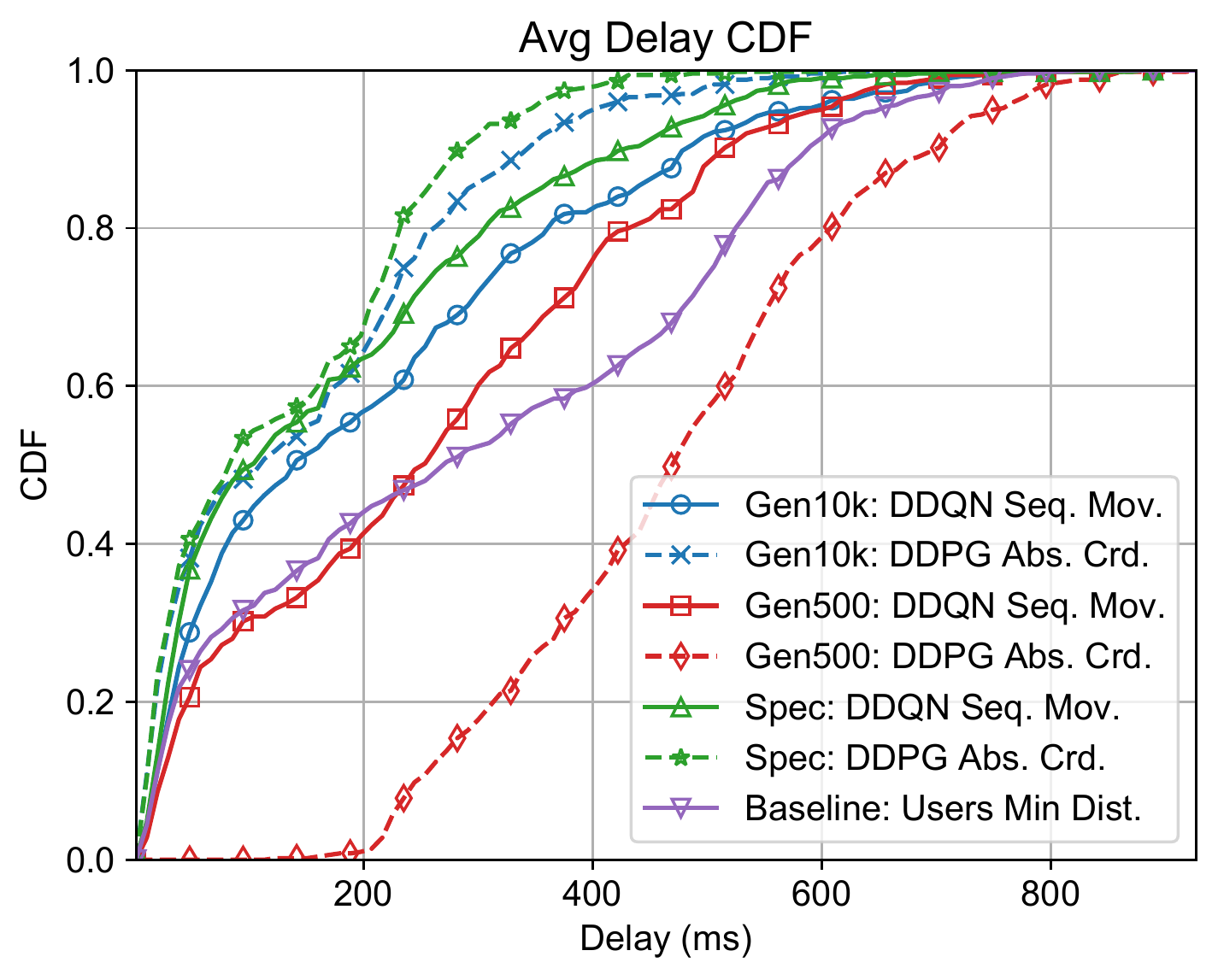}
        \label{Results-Figure: Delay CDF}
    }
    \hfil
    \subfloat[Average PLR CDF.] {
        \includegraphics[width=0.31\linewidth]{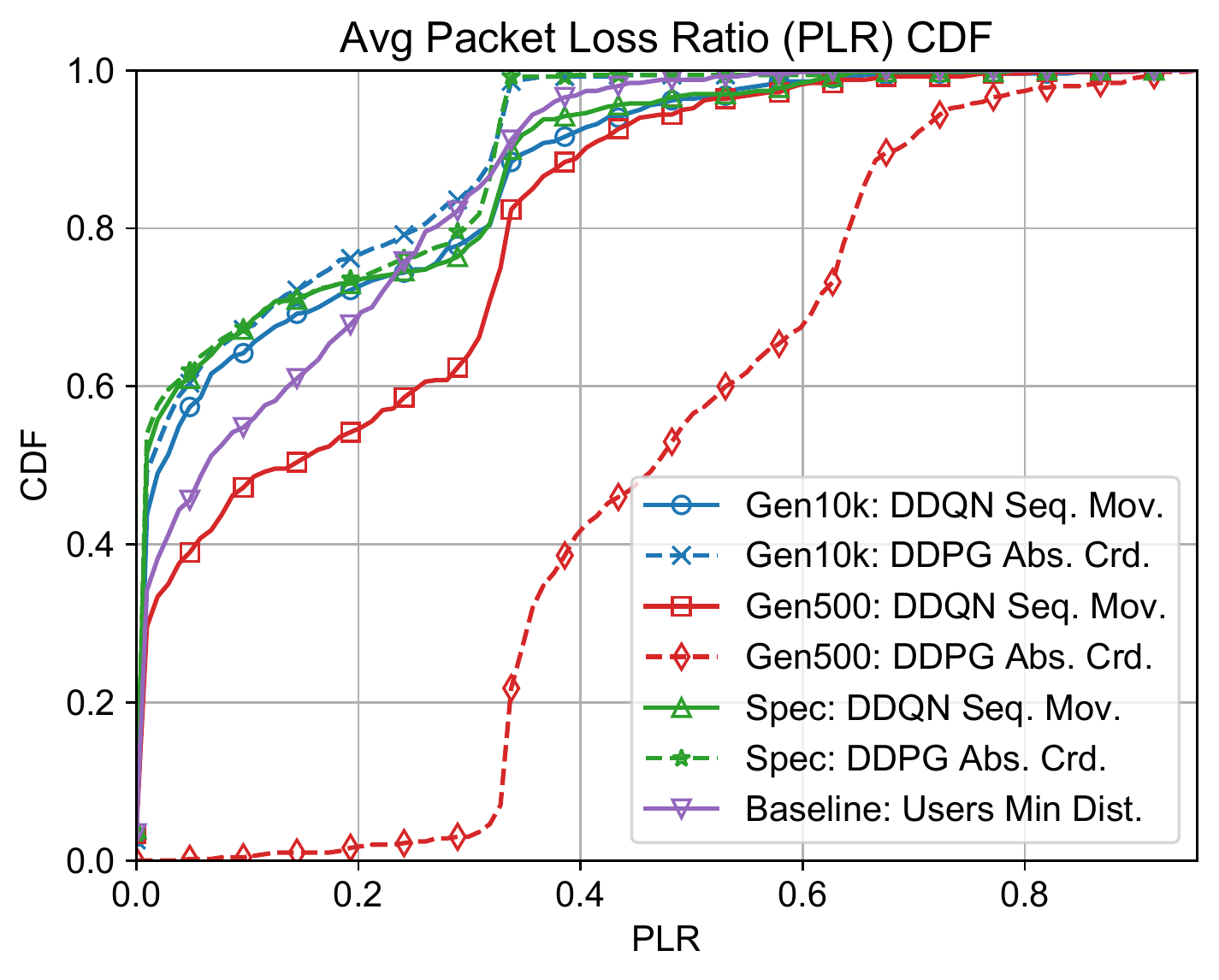}
        \label{Results-Figure: PLR CDF}
    }
    \caption{Average aggregate throughput, delay and PLR CDFs calculated for all test episodes. All versions of TUPA were trained with 10,000 episodes.}
    \label{Results-Figure: Throughput delay PLR CDFs}
\end{figure*}

The Cumulative Distribution Functions (CDFs) of the average network utility, aggregate throughput, delay and PLR calculated for all test scenarios are shown in \cref{Results-Figure: Network utility CDF} and \cref{Results-Figure: Throughput delay PLR CDFs}. The results demonstrate that all versions of TUPA except the \emph{Gen500} provided higher network utility to the users in all scenarios, when compared to the baseline, with a maximum gain of $ \approx $~1.75x for the 25th percentile. Moreover, the average delay demonstrated the greatest improvement relative to the baseline, with a reduction of $ \approx $~0.42x for the 60th percentile.

\begin{figure}
    \centering
    \includegraphics[width=1\linewidth]{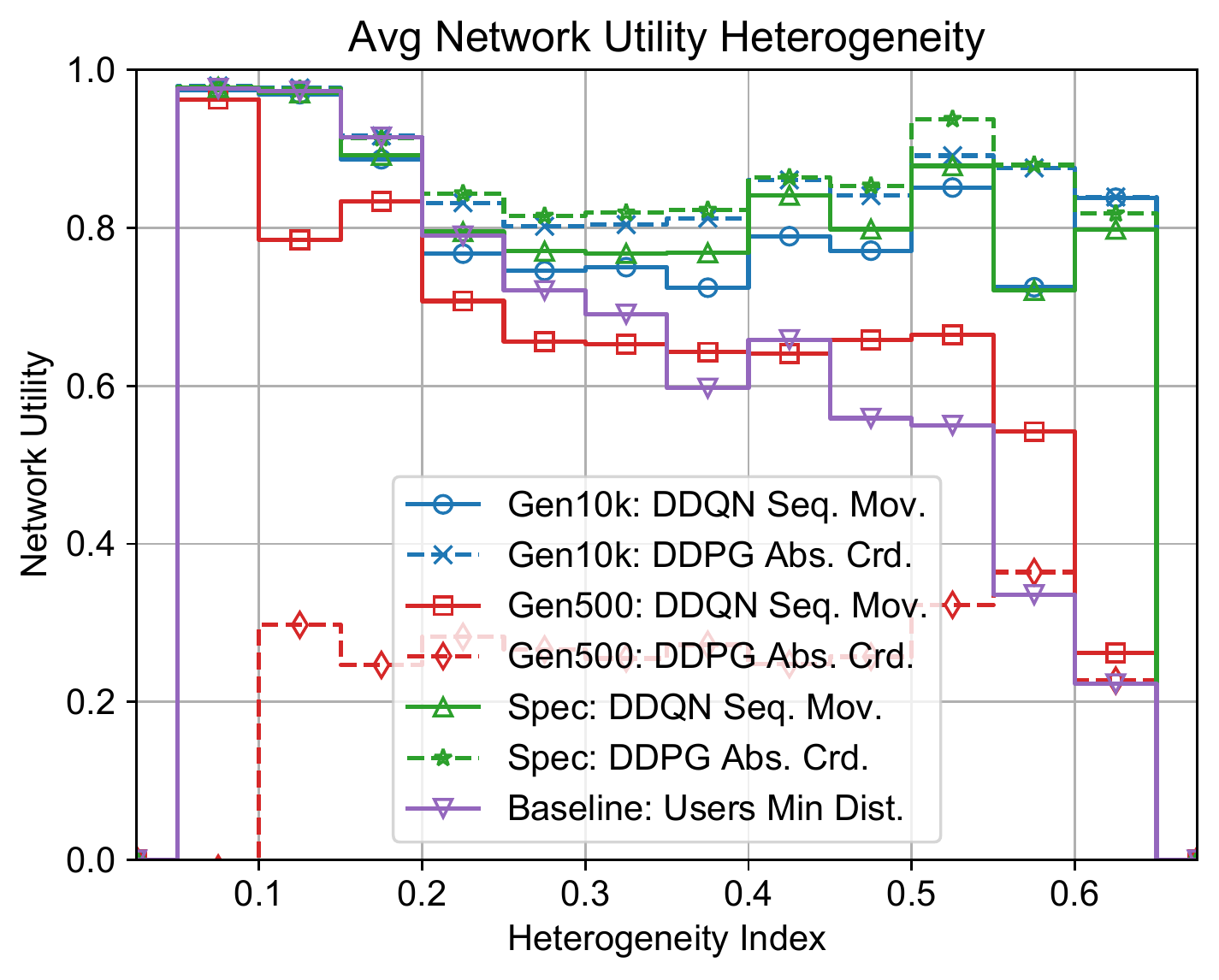}
    \caption{Average network utility per heterogeneity index calculated for all test episodes. All versions of TUPA were trained with 10,000 episodes.}
    \label{Results-Figure: Network utility heterogeneity}
\end{figure}

In order to characterize the performance of TUPA in different types of scenario heterogeneity, we grouped the test scenarios in bins of 0.05 heterogeneity indices and calculated the average network utility of all scenarios in that bin. \cref{Results-Figure: Network utility heterogeneity} shows the average network utility per heterogeneity index bin. It is demonstrated that all versions of TUPA except the \emph{Gen500} increased the average network utility in all bins, with respect to the baseline. Moreover, the results show that the network utility gain relative to the baseline increases with the heterogeneity index. In fact, the most heterogeneous scenarios achieved an average network utility of $ \approx $~0.8, whereas the baseline only achieved $ \approx $~0.2, which corresponds to a gain of $ \approx $~4x.

Furthermore, the results of the two generalization strategies demonstrates that training the FAP agent once with more unique scenarios, as opposed to less unique scenarios during multiple times, enabled it to achieve a similar performance to the specialization strategy.

\subsection{Learning Rate Performance}

To analyze the learning rate of TUPA, we trained it with a different number of training episodes and evaluated its performance after each training session. TUPA was trained with a number of episodes increasing in logarithmic scale from $ 10^2 $ to $ 10^4 $ episodes.

\begin{figure}
    \centering
    \includegraphics[width=1\linewidth]{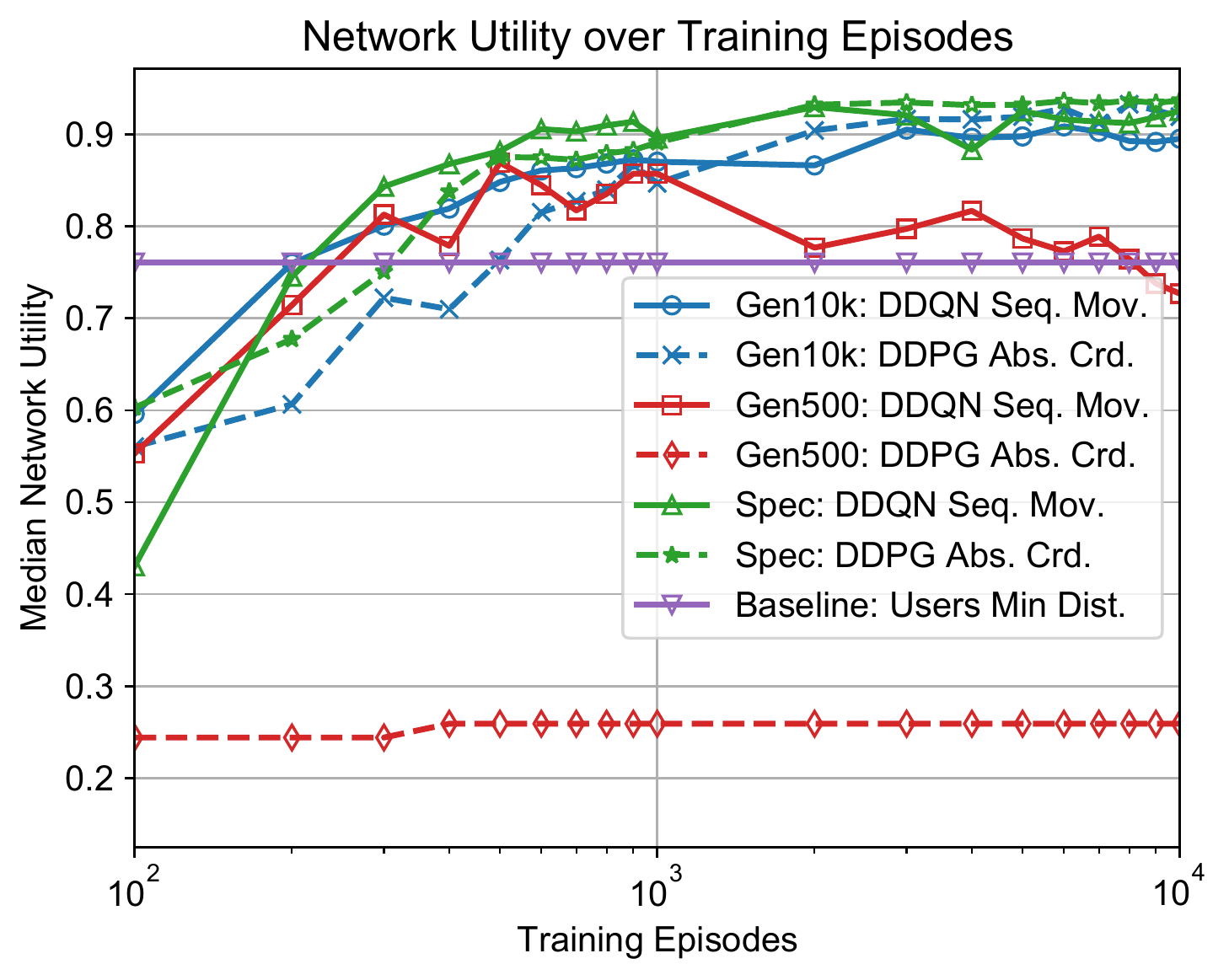}
    \caption{Median value of the average network utility for different numbers of training episodes.}
    \label{Results-Figure: Avg network utility vs. training episodes}
\end{figure}

The results obtained are shown in \cref{Results-Figure: Avg network utility vs. training episodes}. The results show that the median network utility achieved by all versions of TUPA except the \emph{Gen500} increased with the number of training episodes, until a point where it reached the maximum value of $ \approx $~0.9. However, the results of the \emph{Gen500} strategy show that the FAP agent is unable to surpass the performance of the baseline, regardless of the number of training episodes. Additionally, it can be seen that the DDQN algorithms learn faster than the DDPG algorithms. However, the DDPG algorithms achieve slightly better results than the DDQN. Nevertheless, all versions of TUPA except the \emph{Gen500} are able to surpass the network utility performance of the baseline after 500 training episodes, at most. Finally, the results demonstrate that the \emph{Gen10k} generalization strategy achieves similar learning rates as the specialization strategy.

\section{Conclusions} \label{Conclusions-Section}

This paper proposed the traffic-aware UAV placement algorithm TUPA, in order to maximize the network utility provided to the ground users. The results demonstrated that the FAP increased the average network utility compared to the baseline solutions, with a maximum gain of 4x gain in scenarios with heterogeneous traffic demands. Furthermore, the results showed that the proposed DRL training methodology enabled TUPA to generalize the knowledge acquired in the training phase and achieve similar results when applied to unknown variations of those scenarios -- with different combinations of users' positions and traffic demands -- with no additional training. Additionally, training the FAP agent once with more unique scenarios demonstrated better results than training it multiple times with less unique scenarios. As future work, we plan to extend our work to cooperative multi-UAV algorithms.

\section*{Acknowledgments}

This project has received funding from the European Union’s Horizon 2020 research and innovation programme under grant agreement No 833717.
The first author thanks the funding from FCT under the PhD grant PD/BD/113819/2015.

\bibliographystyle{IEEEtran}
\bibliography{References}

\end{document}